%% file: PT_access.tex
\documentclass[acmsmall]{acmart}

\usepackage{soul}
\newcommand{\dcc}{da/cc}
\newcommand{\rev}[1]{{\textcolor{black}{#1}}}
\newcommand{\myrev}[1]{{\textcolor{black}{#1}}}
\usepackage{hyperref}
\usepackage{arydshln}

\AtBeginDocument{%
  \providecommand\BibTeX{{%
    \normalfont B\kern-0.5em{\scshape i\kern-0.25em b}\kern-0.8em\TeX}}}

\setcopyright{acmcopyright}
\copyrightyear{2018}
\acmYear{2018}
\acmDOI{XXXXXXX.XXXXXXX}

\acmConference[Conference acronym 'XX]{Make sure to enter the correct
  conference title from your rights confirmation emai}{June 03--05,
  2018}{Woodstock, NY}
\acmPrice{15.00}
\acmISBN{978-1-4503-XXXX-X/18/06}

\begin{document}

\title{``I'm Just Overwhelmed'': Investigating Physical Therapy Accessibility and Technology Interventions for People with Disabilities and/or Chronic Conditions}
\renewcommand{\shorttitle}{``I'm Just Overwhelmed''}
\author{Momona Yamagami}
\email{my13@uw.edu}
\affiliation{%
  \institution{Department of Electrical \& Computer Engineering, University of Washington, Seattle}
  \streetaddress{185 E Stevens Way NE}
  \city{Seattle}
  \state{Washington}
  \country{USA}
  \postcode{98195-2350}
}

\author{Kelly Mack}
\affiliation{%
  \institution{Department of Computer Science \& Engineering, University of Washington, Seattle}
  \streetaddress{185 E Stevens Way NE}
  \city{Seattle}
  \state{Washington}
  \country{USA}
  \postcode{98195-2350}
}

\author{Jennifer Mankoff}
\affiliation{%
  \institution{Department of Computer Science \& Engineering, University of Washington, Seattle}
  \streetaddress{185 E Stevens Way NE}
  \city{Seattle}
  \state{Washington}
  \country{USA}
  \postcode{98195-2350}
}

\author{Katherine M. Steele}
\affiliation{%
  \institution{Department of Mechanical Engineering, University of Washington, Seattle}
  \streetaddress{3900 E Stevens Way NE}
  \city{Seattle}
  \state{Washington}
  \country{USA}
  \postcode{98195-2350}
}

\renewcommand{\shortauthors}{Yamagami et. al}

\begin{abstract}
    Many individuals with disabilities and/or chronic conditions (\dcc) experience symptoms that may require intermittent or on-going medical care.
    However, healthcare is an often-overlooked domain for accessibility work, where 
    access needs associated with temporary and long-term disability must be addressed to increase the utility of physical and digital interactions with healthcare workers and spaces. 
    Our work focuses on a specific domain of healthcare often used by individuals with \dcc: physical therapy (PT). 
    Through a twelve-person interview study, we examined how people's access to PT for their \dcc~ is hampered by social (e.g., physically visiting a PT clinic) and physiological (e.g., chronic pain) barriers, and how technology could improve PT access.
    In-person PT is often inaccessible to our participants due to lack of transportation and insufficient insurance coverage.
    As such, many of our participants relied on at-home PT to manage their \dcc~ symptoms and work towards PT goals.
    Participants felt that PT barriers, such as having particularly bad symptoms or feeling short on time, could be addressed with well-designed technology that flexibly adapts to the person's dynamically changing needs while supporting their PT goals.
    We introduce core design principles (adaptability, movement tracking, community building) and tensions (insurance) to consider when developing technology to support PT access.
    Rethinking \dcc~ access to PT from a lens that includes social and physiological barriers presents opportunities to integrate accessibility and adaptability into PT technology.
\end{abstract}

\ccsdesc[500]{Human-centered computing~Accessibility}

\keywords{accessibility, physical therapy, disability, user needs, chronic conditions, chronic illness}


\maketitle

\input{PTaccess/sec_intro}
\input{PTaccess/sec_relatedwork}
\input{PTaccess/sec_methods}
\input{PTaccess/sec_results}
\input{PTaccess/sec_discussion}

\section{Conclusion}
Through our interview study with twelve individuals who self-identify as having a disability and/or chronic condition (\dcc), we gained insights into the complex issues that this group of individuals face at the intersection of social and physiological barriers when accessing physical therapy (PT). Our findings characterize both in-person and at-home PT as imperfect solutions due to transportation, insurance, safety, and equipment issues \rev{and identify barriers that technology could help alleviate}. Notably, PT routines that do not center the daily experiences of people with \dcc are inherently challenging and less effective. For example, daily symptom fluctuations can make it difficult or impossible to complete prescribed PT. Through our study, we characterized factors that influence a person's PT experience including motivators and access challenges, which outline a clear set of needs that technology can fill to support PT. Building technology that is motivating and adaptable between and within users and that includes key factors (e.g., tracking progress, data privacy) may support PT accessibility. Finally, our work highlights the tightly interwoven threads of social and physiological barriers to PT access. Our analysis would be incomplete looking at one or the other, and they cannot be disentangled. 
We call for other researchers to thoughtfully consider the nuanced inter-relation between physiological and social access needs in accessibility work moving forward, particularly when studying people with chronic conditions.

\section{Acknowledgments}

We thank all the participants that took the time for the interview.
This research was supported by a grant from the National Institute of Biomedical Imaging and Bioengineering of the National Institutes of Health (R01EB021935) and the Center for Research and Education on Accessible Technology and Experiences (CREATE).




\bibliographystyle{ACM-Reference-Format}
\bibliography{sample}

\end{document}

%% file: PTaccess/sec_intro.tex
\section{Introduction}
Chronic conditions are highly prevalent, with six in ten adults in the US having a chronic condition~\cite{cdcchronic}.
Many individuals with chronic conditions experience \textit{``uncertain, unpredictable, [and potentially] progressively deteriorating illness''}~\cite{baker1989relations} that may require medical care~\cite{pinder1999zones}.
\rev{Yet, prior work around the inaccessibility of medical care~\cite{iezzoni2021physicians,edwards2020barriers,lagu2013access,pharr2019accessibility} and physician bias towards patients with disabilities~\cite{iezzoni2021physicians} raises significant concerns about people with disabilities' equitable access to healthcare. }
Physical therapy (PT) is a particularly important domain in which to investigate healthcare accessibility concerns, given its frequent use by individuals with a variety of chronic conditions that may impact access to PT.
A better understanding of social and physiological\footnote{We refer to the set of factors within a person's body/mind including physical body parts but also psychological/emotional processes.} barriers to PT accessibility can inform the design of technology supports, as well as improve our understanding of the PT service ecosystem.
In this paper, we examine the inter-relationship of managing embodied illness experiences (e.g., physical pain) and fighting socially created, exclusionary accessibility barriers~\cite{cobley2018disability} in the context of PT access. 
We discuss the potential for \rev{ technology to alleviate these barriers~\cite{mankoff2010disability} for people with disabilities and/or chronic conditions (\dcc)\footnote{We specify \dcc~ because many people with chronic conditions do not necessarily identify as having a disability, and vice versa.} and introduce design principles to guide future innovation.}

PT is an important health service for acute care such as after an injury (e.g., ankle sprain, back pain), and as part of long-term care strategies for chronic conditions (e.g., Ehlers-Danlos syndrome, fibromyalgia)~\cite{physical2001}. 
During PT, an individual and their physical therapist(s) work together to identify and achieve goals by assessing and implementing an (often movement-based) intervention~\cite{worldphysiotherapy}. 
These interventions can be prescribed as in-person sessions, where the individual meets with the physical therapist to do exercises together, or as at-home sessions, where the individual is prescribed exercises to do at-home. 
Adherence to the PT exercise routines is important for achieving the individual's goals~\cite{langhorne2009motor,lohse2014more,rutten2010adherence}, but is often low~\cite{forkan2006exercise,melillo1996perceptions,vermeire2001patient}, due to challenges such as not having time for the exercises, daily stress, and lack of social support and guidance~\cite{essery2017predictors}. 
Engagement in PT exercises is one challenge where design recommendations for technology interventions~\cite{mason2019design,alankus2010towards} have been developed, such as collecting at-home compliance and performance data \cite{huang2014technology}
or  gamifying PT exercises with commercially available gaming devices (e.g., Nintendo Wii, virtual reality headsets)~\cite{alankus2015reducing,thomson2014commercial,friedman2014retraining,laver2017virtual,matamala2020role}.
However, many of these technology interventions do not take into account the unpredictable, fluctuating symptoms that people may experience from their \dcc~~ and how those symptoms interact with social barriers to influence PT access in complex ways.
\rev{Moreover, while prior work focuses mostly on either physiological~\cite{huang2014technology} or social~\cite{mason2019design,alankus2010towards} barriers to PT, considering the nuanced interaction between physiological and social barriers is important to support PT access for people with \dcc~ and inform technology design.}
Rethinking \dcc~~ access to PT from this more holistic lens presents opportunities to develop technology to improve PT accessibility.

We present an interview study with twelve people with \dcc~
in the United States  who do PT exercises at-home or with a physical therapist in-person.
Our study examines the interaction of disability with their PT to address three questions:
\begin{enumerate}
    \item What motivates or demotivates people with \dcc~ to do PT? 
    \item What are the social and physiological barriers to PT access for people with a variety of \dcc?
    \item How can technology address these barriers and support people's PT goals?
\end{enumerate}

Our interviews revealed the complexities of having one or multiple \dcc~ and participating in PT. We found that \dcc~ often served as a motivator to engage in PT, for example, to maintain current physical abilities, relieve chronic symptoms, or achieve a physical feat (e.g., engaging in a sport). 
At the same time, people with \dcc~ sometimes found that PT exercises required modification to be accessible, such as when having a ``bad symptoms day''. 
In addition to physiological barriers, our participants discussed accessibility challenges that reflected socially constructed barriers, including trouble finding transportation to get to and from in-person PT and challenges with insurance. 
\rev{Our findings also revealed the complex interactions between physiological and social barriers to PT, such as feeling too sick on the day of an appointment to drive to and perform PT.}
Finally, we present insights from participants about how technology could support at-home PT. Participants were excited for such systems because of the potential to customize daily exercises or converse with a physical therapist outside of the clinic; both are examples of important features when one's abilities fluctuate daily due to a \dcc~~ or if in-person PT is not available to them.

Our contributions include:
\begin{enumerate}
    \item an assessment of lived experiences of people with \dcc~ who are doing PT,
    \item identification of opportunities for technology to alleviate social and physiological barriers to PT, and 
    \item an introduction of design principles from participant technology design suggestions that could improve at-home PT access.
\end{enumerate}

\rev{While technology development is only one aspect of improving accessibility to PT for people with \dcc, our results suggest that there are some PT access barriers that technology is uniquely situated to address, such as dynamically updating exercises or providing opportunities to connect with other people with \dcc~ doing PT.
Considering the complex interplay between social and physiological barriers to PT presents opportunities to develop technology that holistically supports PT accessibility for people with \dcc. 
} 

%% file: PTaccess/sec_relatedwork.tex
\section{Related Work}

There has recently been a shift towards a more nuanced interpretation of disability and impairment, that encourages dismantling social access barriers while acknowledging that for some, their bodies are at the center of their experience with disability~\cite{mankoff2010disability,williams1999anybody,shakespeare2013disability,shakespeare2010beyond}. 
PT sits squarely in this intersection.
\rev{
We identify the post-modern model of disability as a useful framework for understanding the complex interplay between social and physiological barriers. 
Applying this model to PT and healthcare accessibility highlights still underexamined access barriers and opportunities for technology interventions. 
}

\subsection{Defining Disability}

In examining the PT experience for people with \dcc, it is first important to examine how we define disability and impairments. 
Two of the most prevalent models of disability -- the social and the medical model -- have limitations that do not fully address the experiences of people with disabilities. 
Although less prevalent, other models of disability highlight different factors core to the disabled experience, such as interdependence~\cite{bennett2018interdependence} or social effects~\cite{world2007international}. Notably, the post-modern model takes into account both structural/social factors and medical/physiological factors can be helpful to address these limitations ~\cite{mankoff2010disability,williams1999anybody,shakespeare2013disability,shakespeare2010beyond}.
Key to this approach is to consider \textit{impairment} (i.e., physical or mental differences or limitations) and \textit{disability} (i.e., social factors that limit a person's participation due to differences in ability) as interrelated in nuanced ways ~\cite{cobley2018disability,shakespeare2010beyond}.  

The medical and social models of disability differ in describing what ``causes'' the disabling experiences. The \textit{medical model} has often been used in the medical or assistive technology fields because it \textit{“focuses on the physical and functional limitations a person may demonstrate”~}\cite{mankoff2010disability}. 
In contrast, a traditional interpretation of the \textit{social model} frames disability as arising from mismatches between a person's ability and the world. 
When people with disabilities face an access challenge, the blame for inaccessibility falls not on the individual, but on society's  laws, architecture, and ableist enforcers of exclusionary, discriminatory  practices in life, work, and education ~\cite{cobley2018disability,williams1999anybody,shakespeare2013disability,shakespeare2010beyond}.
However, a pure social model interpretation does not recognize the tight relation between disability and the body~\cite{clare2001stolen}. 
Wendell argues: \textit{``some unhealthy disabled people, as well as some healthy people with disabilities, experience physical or psychological burdens that no amount of social justice can eliminate''}~\cite{wendell2001unhealthy}.

Alternative models of disability aim to address the tension between the social and medical models (e.g., \cite{bennett2018interdependence,world2007international,mankoff2010disability}). Among these, the interdependence model~\cite{bennett2018interdependence} can be a useful lens to view medical care, and physical therapists in particular, as part of an interdependent network that supports individuals across the lifespan.  
In medicine, the International Classification of Functioning, Disability and Health (ICF) model~\cite{world2007international} has been used more heavily in recent years to understand the environmental and social barriers that limit function in addition to physiological barriers.
In HCI, the post-modern model of disability \textit{``privileges each individual's unique lived experience...disability, illness, impairment, functional limitation, and bodily anomaly are separate but complementary issues, and successful assistive technology must account for all of these perspectives''}~\cite{mankoff2010disability}. 
While an interdependence or ICF model highlight key factors in disability, a post-modern model emphasizes the interplay between social and physiological barriers.

Thus, the post-modern model of disability is a useful perspective to adopt in analyzing PT, a system which is socially constructed and maintained through processes like insurance, but one that is focused deeply on quality of physical movement.
Indeed, the disabling situations, both created by an ableist society and physical differences in bodies, have important roles to play in understanding someone's experience with disability~\cite{pinder1996sick,baker1989relations}.
Particularly for people who may be involved in PT, not all daily pain and discomfort can be assuaged solely by a social interpretation of accessible practices. 
On the other hand, adopting the pervasive, and harmful assumption that physical and functional limitations are problematic and must be overcome by technology also does people a disservice. 
%
\rev{Instead of making solely social or physiologically based assumptions about why PT is inaccessible, the post-modern model of disability provides a lens to examine each individual's experience of the complex interplay between social and physiological access barriers.}


\subsection{Defining Physical Therapy}

The primary goals of doing PT are defined by the person doing PT and often include  improving or maintaining movement and function. 
Physical therapists and the individual doing PT strive to achieve this through a feedback loop of evaluation, plan development, implementation, and assessment ~\cite{kisner2017therapeutic}. 
The development of the plan of care generally involves setting measurable short- and long-term outcome goals and considering the resources available to the person. 
During the course of treatment, people generally meet with their physical therapists anywhere from weekly to monthly.
People are encouraged to follow a (usually daily) PT exercise program that is tailored towards them and their goals to supplement the in-person PT sessions.
When doing PT to improve symptoms of \dcc, high intensity or dosage can be important for recovery and to maintain quality of life~\cite{lohse2014more}. For example, the Levine protocol~\cite{shibata2012short} is a rigorous eight-month cardiovascular and strength training routine for people with postural orthostatic tachycardia syndrome (POTS) where people work up to exercising 5-6 hours per week. The exercises, frequency of the exercises, duration, and number of repetitions are usually provided to people on a sheet of paper, and adapted during the in-person PT sessions as people progress.

\rev{However, many accessibility barriers exist throughout the PT process that makes completing a treatment plan less accessible, or even hinder starting PT in the first place. }
\rev{Technology has been developed to address a limited number of these barriers, but few studies provide guidelines or develop technology to address the intersectional needs of someone with \dcc~ whose symptoms may fluctuate. We detail both of these areas below.}


\subsubsection{\rev{Barriers to PT}}
Access and adherence to in-person and at-home PT can be challenging for numerous reasons. For example, it can be challenging to obtain the in-PT dosage required for a \dcc~ due to the prohibitive cost and limited number of PT appointments a person can make with insurance~\cite{matamala2020role,shaw2008hospital, deyle2005physical}.
In rural and/or low-and-middle income areas, lack of experienced physical therapists, insufficient community-based programs, long travel times, and high cost to travel are also significant barriers to attending in-person PT~\cite{chumbler2015randomized,sarfo2017potential,world2010telemedicine}.
Unfortunately, adherence to at-home PT programs also comes with its own challenges such as limited time, daily stress, and lack of social support and guidance ~\cite{bassett2003assessment,essery2017predictors}.
%
\rev{There is room for innovation in the space of PT by adapting a new perspective around adherence failure. Lack of adherence is often viewed as a failure of the person doing PT to comply with medical advice~\cite{essery2017predictors}.
A better understanding of PT success from an accessibility perspective is needed. For example, people with chronic conditions may have strong, uncomfortable symptoms that may worsen when doing exercises or symptoms that vary day-to-day~\cite{holland2018whenever,pinder1996sick,priestley1999unfinished,shaw2008hospital}. 
This variability within (e.g., daily fluctuation) and between people with chronic conditions is understudied, or viewed simply as an adherence issue, rather than an opportunity for improved technology design to help manage this variability ~\cite{mack2021what}. 
}

\subsubsection{Existing Technology to Facilitate PT Access}

\rev{
Technology is well-suited to address some PT access barriers, such as 
reminding people to do the PT exercises~\cite{micallef2016time,holden2015cues}, facilitating data sharing between the person doing PT and the physical therapist~\cite{malu2017sharing}, tracking at-home PT for quantity and quality~\cite{lee2019learning,huang2014technology}, and gamification to enhance PT exercise engagement~\cite{thomson2014commercial,friedman2014retraining, alankus2011stroke,alankus2015reducing,mason2019design,cheng2015towards,alankus2010towards}. 
Wrist-worn devices that remind people to do their PT exercises through visuals, sound, or vibration improves people's activity levels~\cite{holden2015cues} and may help people stick with an exercise routine~\cite{micallef2016time}.
An online platform to share PT exercises and progress with one's physical therapist can improve motivation and provide a sense of community~\cite{malu2017sharing}.
Assessing at-home PT quality can also help people get feedback on their movement quality when a physical therapist is not available~\cite{lee2019learning} and help track progress and improve motivation to complete their PT routine~\cite{huang2014technology}. 
Similarly, gamification of PT exercises by mapping exercises to movements of an avatar on a computer screen can help improve adherence and engagement with PT~\cite{thomson2014commercial,friedman2014retraining, cheng2015towards} and encourage full range of motion ~\cite{alankus2010towards,alankus2011stroke,alankus2015reducing}.
}

\rev{While such technologies help alleviate specific physiological or social PT barriers, further research is needed on how technology can holistically support PT accessibility for people with \dcc.
Many of these technologies focus on improving adherence to PT exercise routines with the assumption that the exercises are routinely evaluated and updated by a physical therapist \myrev{and that the physical therapist has a holistic understanding of the needs of the individual doing PT}. 
However, consistent access to in-person PT is often a major obstacle to PT accessibility, \myrev{and this obstacle is often compounded by social barriers like ableism or medical bias~\cite{kirschner2009educating,iezzoni2021physicians}.
This could result in the physical therapist prescribing exercises that are not appropriate or accessible~\cite{rinne2016democratizing}, requiring clients to advocate for themselves or adapt exercises on their own.}
An understanding of how technology can support PT accessibility when taking into account the possibility that people may be doing at-home PT without the support of a physical therapist while experiencing fluctuating symptoms due to their \dcc~  is largely missing.
Social and physiological PT barriers are often researched separately, but their effects are challenging to disentangle; both must be considered holistically during technology development.
}

%% file: PTaccess/sec_methods.tex
\section{Methods}

We performed semi-structured interviews with twelve U.S.-based, \dcc~  individuals who are currently doing PT either with a physical therapist and/or on their own at home. 
\myrev{Our specific inclusion criteria were broad to recruit a diverse sample, including: 1) between the ages of 18-90; 2) have a disability or chronic condition; 3) currently do physical therapy with a physical therapist or at-home for their \dcc. Our exclusion criteria were 1) do not have access to a computer or phone that can be used to call into a remote interview; 2) not able to provide consent independently.
Participants were initially recruited from social media posts on Facebook Twitter, and Reddit on communities focused on disability and accessibility to recruit people who attended PT for a variety of reasons and diversity in abilities. We then used snowball sampling until we achieved saturation.
}
Participants filled out a screening survey with demographic information and we selected participants to include multiple disabilities, genders, and races. The interviews were one hour-long on a video calling platform, due to COVID-19 restrictions. All participants were compensated with a \$15 Amazon gift card, and the interviews were recorded and transcribed by hand. The protocol was approved by our Institutional Review Board.

\subsection{Interview Protocol}

Our interviews were composed of three main parts \myrev{(see supplemental for sample interview questions)}. First, we discussed the participant's background with technology and disability. We discussed the technologies that the participants owned to understand potential sensors to use for PT tracking. We then asked about the participant's disability, including when they became disabled and how their disability affects their daily life. We also recorded what assistive technologies they used \myrev{to perform activities of daily living including mobility tools and adaptive tools to improve device accessibility.
We let participants decide what were assistive technologies to them during the interview.
During analysis, we considered any digital (e.g., screenreader) or physical (e.g., wheelchair) device or tool used to make activities of daily living easier as an assistive technology.}
\myrev{We also asked about any} accessibility challenges they faced with their devices.

In the second part of the interview, we asked the participant about their experience with PT, including the exercises they perform, their motivation for doing PT, and their goals that they are working towards. 
We asked them to differentiate their PT routines done with professional physical therapists (if they saw one) and routines they performed at home. 
We also discussed what factors demotivated them to do PT and the effect that their \dcc~ had on their exercises. 

In the third part of the interview, we  asked participants about potential sensor-based systems to aid in at-home PT. 
We provided examples like using accelerometers to track a PT exercise that then enables technology rewards like playing an episode of Netflix or opening social media. 
\myrev{The provided examples were inspired by current existing and/or popular technologies to support physical activity in general, such as wearable activity monitors (e.g., Fitbit)~\cite{brickwood2019consumer}, mobile phone interventions (e.g., text reminders), and active video games or exergaming via mobile phone (e.g., Pokemon Go)~\cite{king2019physical} or game console (e.g., Nintendo Wii)~\cite{kappen2019older}.
}
We talked with participants about when they would want the system to prompt them to do exercises, what exercises they would be willing to do, and how they could dismiss the notifications to do exercises. 
Given our imagined form factor involving smart devices, we asked if there were unique ways that our system could motivate them to complete exercises (e.g., the user must do 10 squats before they can open Instagram), and if they would want the exercises to be logged manually or tracked automatically via sensor data.

We concluded the interview with demographic questions and information about compensation.

\subsection{Analysis}

Interview transcripts were analyzed using reflexive thematic coding \cite{braun2006using, braun2019reflecting}. 
One author conducted all of the interviews and reviewed all of the interview transcripts to take notes and develop codes. Another author performed the same process on interview segments to check for gaps or biases in code coverage. The full list of 233 codes were discussed and revised by the first two authors.
After the code book was complete, one author applied the final codes to all of the transcripts. 
After all the data was coded, the authors met and through discussion arrived at the broader themes presented in the results. 

\subsection{Positionality Statement}

We recognize that, in performing reflexive thematic analysis, our backgrounds and positionality shaped our findings. This work was conducted by a mix of disabled and non-disabled white and Asian scholars who work in US institutions. Of the four authors, two are engineers who predominantly do research in improving rehabilitation outcomes. The other two authors are computer scientists engaged in mostly accessibility research. Three of the authors have experience receiving PT for a chronic condition and one author has supported multiple family members through receiving PT for chronic conditions, both via telehealth and in-person, as well as fighting for insurance supported access to  PT for a chronic condition.

\subsection{Participants}

\begin{table}[h]
\caption{The self-defined \dcc~ of each participant.}~\label{tab:participants}
\Description{Each of the 12 participants' self-defined disabilities and/or chronic condition(s).}
\begin{tabular}{l:l|l:l}
\textbf{ID} & \textbf{Disability} & \textbf{ID} & \textbf{Disability} 
\\ \hline
P01 & Spinal injury affecting lower body & P07 & Vestibular neuritis, depression \\ \hline
P02 & \begin{tabular}[c]{@{}l@{}}Lower-body left-side \\hemiparesis, learning disability\end{tabular} & P08 & \begin{tabular}[c]{@{}l@{}}Gastroporesis, anxiety, depression, \\ chronic back pain, torn miniscus\end{tabular} \\ \hline
P03 & Epidermolysis bullosa, acid reflux & P09 & Herniated disc, radiculopathy \\ \hline
P04 & \begin{tabular}[c]{@{}l@{}}Carpal tunnel, herniated disc, \\ hydrocephalus, epilepsy, \\ inverted scoliosis, fibromyalgia\end{tabular} & P10                     & \begin{tabular}[c]{@{}l@{}}Autism, Ehlers-Danlos, \\ chronic pain, complex PTSD, \\ anxiety, depression, dyslexia\end{tabular}                     \\ \hline
P05 & \begin{tabular}[c]{@{}l@{}}Pulmonary lymphoma, irritable\\ bowel syndrome, bronchospasm \end{tabular} & P11 & \begin{tabular}[c]{@{}l@{}}Hypermobile Ehlers-Danlos,  \\partially deaf in one ear\end{tabular} \\ \hline
P06 & Chronic ankle pain after surgery & P12 & \begin{tabular}[c]{@{}l@{}}Ehlers-Danlos, autoimmune disease, \\ POTS, orthostatis hypotension, \\ celiac disease, chronic gastoritis\end{tabular}\\
\hline
\end{tabular}
\newline\newline
\small
Disabilities included autism, cognitive disabilities, learning disabilities, mental health disabilities, health-related disabilities, and motor disabilities. Ten out of twelve participants had multiple \dcc.

Abbreviations: PTSD = post-traumatic stress disorder; POTS = postural tachycardia syndrome
\end{table}

In total, we recruited and interviewed twelve participants from December 2020 - March 2021. 
Six participants identified as men, five as women, and one as non-binary. The participants' races included Caucasian/white (9), white Hispanic (1), African American (1), and Asian (1). The mean age was 32.4 (range = 20-58). The highest degrees earned were high school diploma (3), Bachelor's degree (3), Master's degree (3), Associate degree (1), and other graduate degree (2). 
The \dcc~of each individual are listed in \autoref{tab:participants}, and
we briefly summarize below the participants' self-reported PT background and how their \dcc~ affects technology accessibility and daily life.

\myrev{Participants performed PT exercises at home daily (11), only when needed (2), and/or as much as possible (2). 
The exercises included lower body mobility exercises (8) like walking, running, and biking, stretching or doing yoga (7), lower (6) and upper (4) body resistance band training, and core exercises (4). 
For participants who were prescribed PT exercises by a physical therapist to do at home, many participants felt that they did less PT than they were prescribed (8).} 

The participants' \dcc~affected the accessibility of activities such as work and employment (5),
socialization with family and friends (4),
and keeping up with non-disabled peers (2).
Access barriers stemmed from \dcc~ symptoms such as chronic pain (6),
fatigue (3), 
and nausea (1), 
their \dcc~ making it hard to sit, stand, or walk (6), 
and the effect of their \dcc~ on their mental health (3). 
\rev{Technology adaptations like speech to text (5), ergonomic keyboards and mice (5), and mobility devices (5) were helpful in improving access to activities.}

Further, the \dcc~that our participants had affected their technology use. Seven participants
discussed challenges physically interacting with devices, such as hand fatigue (5),
interacting with a standard mouse or keyboard (4),
using a touchscreen, trackpad, or fingerprint detector (2),
and pressing multiple buttons at once (4).
Consuming digital content caused similar challenges. Six participants
discussed how technology use exacerbates \dcc~ symptoms, including pain, strain, headaches, and motion sickness.
Consequently, participants used a variety of low- and high-tech assistive technologies ranging from screen readers, to ergonomic devices, to placing adhesive tape on devices to allow for better grips.
\rev{Accessibility barriers meant that five participants had to abstain from using technology to alleviate \dcc~ symptoms; these cases demonstrate the importance of designing PT-focused technologies with accessibility in mind.}

%% file: PTaccess/sec_results.tex
\section{Results}

Our participants described many benefits and motivations for doing PT; however, in doing so, they encountered access barriers that were based in the embodiment of participant impairments, social constraints like financial and physical access to PT, and interactions between barriers. We then discuss how this information around motivators and challenges influenced participant's visions for future technologies supporting PT. 
\subsection{\rev{PT Improving Quality of Life is Motivating}}
Many people were motivated to do PT because of its potential to make areas of life more accessible or comfortable given their \dcc, such as working, socializing, and exercising.

For some participants, aiming to close the performance gap between their current abilities and those of the peers or themselves pre-\dcc~was motivational: 
\textit{``I know what it's like to be healthy and strong and I want that again''} (P10). 
%
For others, their motivation stemmed from maintaining quality of life.
P02 discussed: \textit{``my function is deteriorating, so I want to maintain my function and my current level of mobility as long as possible''} and P05 was motivated \textit{``to slow down the progression of disease''}.
Improving and maintaining performance are different goals that affect what ``progress'' looks like and can impact how PT may be presented with technology. 

Ten participants discussed improving mobility and strength as a tangible goal for maintaining or regaining quality of life (P01-P04, P06, P08-P12). 
\textit{``Increasing mobility and the amount that I can walk'' }(P02) was a goal for P01-P03, P06, P10, and P12. P01 describes how PT could increase their physical abilities, allowing them more freedom from social barriers: \textit{``so I would be able to walk around a store or go outside without having to bring my wheelchair...it's kind of a pain because I have to make sure the buildings that I go into are ramp accessible before I go there'' } (P01), as many places are not accessible for people who use mobility devices.
Gaining strength and stamina was also an important goal for P02, P09, P10 and P12 so they can participate in different activities like \textit{``hiking [and] kayaking''} (P12). These smaller, more actionable steps that support higher level goals provide opportunities for technology to support tracking and celebrating progress, which may increase participants' motivation.

Participants also faced chronic symptoms due to their \dcc, which lowered their quality of life and made activities less comfortable.
PT served as a way to combat these negative symptoms, such as pain (P02, P04-P06, P08, P09, P11) or dizziness (P07). 
Similarly, P03, P04 and P12 felt that, by improving their strength and stamina, they could avoid symptoms of their \dcc~ such as pain and scar tissue buildup. Two participants (P02, P11) performed PT intermittently and used their pain levels as a gauge for when to restart PT exercises.

Overall, we saw that participants had clear motivations for engaging in PT, which usually revolved around maintaining or increasing physical capability, or increasing quality of life through symptom mitigation. 
\rev{A purely medical goal for PT exercise compliance is completing the number of exercises set by the physical therapist; 
framing this goal in the context of their higher level goal (e.g., maintaining physical ability) could be motivating and improve adherence to PT routines~\cite{huang2014technology}. }

\subsection{Context-Dependent Barriers to PT }

In-person (i.e., in clinics) and at-home are two common contexts for performing PT, each of which has their own benefits and challenges. While in person PT can be more engaging and allows access to therapist expertise, it can be physically and monetarily inaccessible. At-home PT, while less expensive, makes it more challenging to stay engaged and safely adapt and  correctly perform exercises based on their current \dcc~ symptoms.

\subsubsection{Barriers to In-Person PT}
In-person PT was inaccessible to many of our participants due to insufficient insurance coverage and lack of transportation.
Insurance issues, such as a gap in coverage (P10, P12), insurance only covering a limited number of sessions per year (P03, P09, P12), high out-of-pocket cost (P11), and being too far out from the initial diagnosis to get PT covered by insurance (P02) were all noted as barriers to in-person PT.
The current insurance system in the US only allows for a certain number of PT visits with a physical therapist per year, and consequently, people often end up doing more at-home PT between or in lieu of in-person visits.
Such financial barriers can be viewed as accessibility barriers both in the literal sense (since they remove access to PT) and also because people with disabilities are more likely to be living in poverty \cite{brucker2015more}. As a result of theses insurance challenges, participants were forced to choose between going to unaffordable PT or exacerbating the symptoms of their \dcc. 
P11 discusses, \textit{``as a way to try to save me money [P11's physical therapist] said let's try to meet less often.''}
Transportation was another commonly noted challenge of in-person PT. Not all of our participants owned their own form of transportation. 
Public transportation is not always available, or not an option during a pandemic. Relying on relatives or friends can also cause additional tensions (P01).
The pandemic further demotivated participants from attending in-person PT due to potential exposures (P04, P09, P11).
For example, P11 discussed how meeting with their PT in-person is \textit{``a calculated risk''} due to their \dcc~ making them high-risk for COVID-19. 

\subsubsection{Barriers to At-Home PT}
Because of the many social barriers to in-person PT, many participants relied on at-home PT exercises to alleviate \dcc~ symptoms and improve mobility and strength.
However, at-home PT also came with significant barriers such as requiring time and effort, causing discomfort and fatigue, dealing with fluctuating symptoms, managing safety, and feeling a lack of engagement and variety. 
Nine participants discussed how doing the PT exercises took effort, time, and caused discomfort (P02-P04, P06-P08, P10-P12). 
For example, P07 discussed how their PT exercises exacerbates their \dcc~ symptoms, discouraging them and making the rest of their day less comfortable. 
Symptoms of \dcc~ can also limit the overall number of activities that a person can do in a day,
making it even more challenging for people to incorporate PT into their everyday routines.
For example, P11 discussed how they were \textit{``just piled on with more and more exercises, it is getting harder to stay on track...it's a challenge being motivated, it's a challenge dealing with the fatigue [one of their \dcc~ symptoms], I've been feeling overwhelmed by it all. I just feel like it's an endless list of things to do''}.
In P11's broader narrative, we see that PT was a time-burden that was compounded by the time to seek other medical care, which is an experience specific to people with illness-based \dcc~\cite{wendell2001unhealthy}.
\rev{P11 additionally had multiple disabilities that they went to multiple physical therapists for, who all gave PT exercises that were solely specific to the problem at hand, without considering the larger picture of P11's multiple conditions. }
\rev{P11's narrative highlights how physiological barriers (e.g., fatigue due to \dcc) can interact with social barriers (e.g., navigating the complex healthcare system to find and receive care for multiple conditions), which made it challenging to fit PT exercises into their life.}
\myrev{P11 discussed how their method of coping with this challenge was to \textit{``try to do something at least every day''} (P11).}

\subsubsection{Physiological Barriers to At-Home PT}
Physiological barriers to PT access also affected participants' confidence and motivation to do PT. 
For example, several participants reported  physically not being able to do the PT exercises or not progressing in their PT (P01-P03, P08).
P02 discussed how \textit{``it's very discouraging trying to use the theraband [a resistance band used to do their lower-body PT exercise] sometimes I can't even do it without the theraband resistance...that is one reason why I do those ones less.''}
\myrev{To alleviate the challenge of the prescribed PT exercise not matching how P02 was feeling physically, they self-prescribed other ways to stay active such as \textit{``biking or going on walks...[so] my muscles were still contracting and I was still being active''} (P02).}
Other health symptoms not related to the \dcc~ that the individual is doing PT for could also affect people's motivation. 
For example, P10 discusses how: \textit{``if you're really depressed, getting into the whole, `what's the point of [doing PT], why bother'''} (P10). 
During in-person PT, many of these physiological PT barriers can be mitigated with the help of the physical therapist because they can dynamically adapt the exercises. 
Adapting the PT exercises when doing at-home PT is challenging without access to this expertise.
Many of our participants were given sheets of paper describing the PT exercises, the number of repetitions, and how often to do the PT exercises, which can make it difficult to track and adapt the exercises to suit the person's needs and symptoms. 
\rev{The challenges adapting PT exercises to account for fluctuating symptoms can be considered as both a physiological barrier stemming from their \dcc~ symptoms as well as a social barrier of not having an accessible PT routine that can be easily adapted when people are having a bad symptoms day. }

\subsubsection{Safety Barriers to At-Home PT}
Another challenge of at-home PT was getting feedback on whether participants were doing the PT exercises safely and correctly to prevent injury.
Seven participants (P01, P04, P06, P07, P09-P11) discussed the importance of checking in with their physical therapists when at-home PT exercises didn't alleviate \dcc~ symptoms or if they were not making progress. 
If people were not currently doing in-person PT or they needed feedback at that moment, they resorted to \textit{``comparing myself as best as I can to the visuals on the screen''} (P12) using YouTube videos that they had searched.
However, this can be a challenge because \textit{``you have to be careful and make sure you're not looking at something that's not a good source of information''} (P11). 
Additionally, it is hard to ensure safety when doing at-home PT. 
For example, P07 discussed how falls were a safety concern for some of their PT exercises so they could only do them \textit{``if there’s someone home.''}
To alleviate these challenges, P10 wished they had \textit{``an option to communicate [virtually and asynchronously] with my physical therapist...having that check-in monitoring and what to do next''}. Such a system could mitigate both safety and exercise progression issues.

\subsubsection{Engagement Barriers to At-Home PT}
Lack of engagement and variety in exercises (P02, P05, P09, P12) was another barrier to at-home PT. 
P02 wished the PT exercises \textit{``were more engaging...[the physical therapists] used to try when I was a kid to make them more engaging, like stand on one leg and throw a ball back and forth. But when you get older, [making the PT exercises engaging] is not a thing anymore.''}
Lack of variety in exercises over time exacerbated the disinterest in completing at-home PT exercises. 
P12 also commented that performing exercises in a PT clinic was more engaging than at home because \textit{``at least physically going into PT, there were people you could talk to, you had different tools that we're using, now it's just kind of the same old, same old''}.
P12 alludes to the fact that the PT clinic was a source of socialization, or perhaps even community, which was key to their motivation for PT. 
Therefore, in addition to considering how to make people with \dcc~feel safe, technology that supports at-home PT must address the issues of engagement and community.

Although in-person PT supplemented with at-home PT is considered the norm for exercise-based care for \dcc, our participants discussed how in-person PT is often physically and monetarily inaccessible for them, forcing them to solely rely on at-home PT without physical therapist supervision to manage the symptoms of their \dcc.
While at-home PT removes some barriers like transportation and risk of exposure to illness, it introduces new access barriers that are specific to people with \dcc~ like concerns with safety when doing the exercises. Given the lack of communication channels with physical therapists outside of clinic visits, people with \dcc~ often need to choose which set of barriers (those associated with at-home or in-person) are more appropriate, given their unique situation. 

\subsection{Participant Technology \& Design Suggestions}
All twelve participants were interested in incorporating technology into their at-home PT routines to alleviate at-home PT barriers such as time and effort, fluctuating symptoms, lack of motivation, and safety and injury (Table~\ref{tab:designProbe}). 
Additionally participants cited insurance as a potential concern for several features.
The participants suggested seven technology features that would improve at-home PT access: exercise presentation, tracking, technology rewards, notifications, hardware preferences, data security, and sharing progress.
\myrev{In line with other qualitative work, these suggestions can help inform technology development, but the technology should be tested prior to full deployment.}

\begin{table}[h]
\caption{
Summary of participant suggested features to include when developing technology to support at-home PT and access barriers that are addressed with the features.}~\label{tab:designProbe} 
\Description{From left to right: feature that participants suggested for developing technology to support at-home PT, the alleviated access barrier associated with the feature, a summary of participant suggestions, and a sample quote that highlights why the participants feel the feature is important.}
\begin{tabular}{l:l:l:l}
 \textbf{Feature} & \textbf{\begin{tabular}[c]{@{}l@{}}Access\\Barriers\end{tabular}} & \textbf{Participant Suggestions} & \textbf{Sample Quote} \\ \hline
\textbf{\begin{tabular}[c]{@{}l@{}}Exercise \\ Presentation\end{tabular}} & \begin{tabular}[c]{@{}l@{}}TE \\ FS\end{tabular} & \begin{tabular}[c]{@{}l@{}}Choose between different \\exercises; customize depending \\on day; customize number of\\ repetitions; support exercise \\progression; minimal setup (e.g., \\sit-to-stand)\end{tabular} & \textit{\begin{tabular}[c]{@{}l@{}}``[being able to choose exercises \\is helpful] because certain \\exercises are targeted to help \\with certain areas...so only I\\ would know what I need \\for the day'' (P04)\end{tabular}}\\ \hline
\textbf{Tracking} & \begin{tabular}[c]{@{}l@{}}FS \\ MO\\SI\end{tabular} &  \begin{tabular}[c]{@{}l@{}}Tracking PT progress is \\motivating; feedback useful to \\prevent injury or worsening \\\dcc~ symptoms\end{tabular} & \textit{\begin{tabular}[c]{@{}l@{}}``[I want to] make sure that I'm \\doing [the PT exercises] \\right because if I don't, it's\\ gonna mess [up the ankle]\\ even more'' (P06)\end{tabular}} \\ \hline
\textbf{\begin{tabular}[c]{@{}l@{}}Technology \\ Rewards\end{tabular}} & MO & \begin{tabular}[c]{@{}l@{}}Non-essential apps (e.g., not \\email); time sinks (e.g., social \\media); unlocking devices;\\ commercial breaks (e.g., Netflix); \\congratulatory or fun elements\end{tabular} & \textit{\begin{tabular}[c]{@{}l@{}}``If I'm really in the thick of it \\with Grey's Anatomy, that \\would absolutely motivate \\me [to do PT]'' (P12)\end{tabular}} \\ \hline
\textbf{\begin{tabular}[c]{@{}l@{}}Notifications \\ Timing\end{tabular}} & \begin{tabular}[c]{@{}l@{}}TE \\ FS\\SI\end{tabular} & \begin{tabular}[c]{@{}l@{}}Personalized notification timing\\ and frequency; preemptive \\notifications for \dcc~ symptoms;\\ remind to do later; offer easier \\exercises before dismissing\end{tabular} & \textit{\begin{tabular}[c]{@{}l@{}}``I want it to detect my pain \\somehow and then prompt \\me [to do the PT exercise]''\\ (P09)\end{tabular}}  \\ \hline
\textbf{\begin{tabular}[c]{@{}l@{}}Hardware \\ Preferences\end{tabular}} & \begin{tabular}[c]{@{}l@{}}IN \\ SI \end{tabular} & \begin{tabular}[c]{@{}l@{}}Cameras with depth perception; \\wearables; cost and sensor\\ failure concerns; some preferred\\ software only\end{tabular} & \textit{\begin{tabular}[c]{@{}l@{}}``[software] would make it \\easier to get updates...\\hardware would be a never-\\ending investment'' (P08)\end{tabular}} \\ \hline
\textbf{\begin{tabular}[c]{@{}l@{}}Data \\ Security\end{tabular}} & IN & \begin{tabular}[c]{@{}l@{}}No data security concerns; want \\data to be anonymized; \\HIPAA compliant; data security \\against insurance is concerning\end{tabular} & \textit{\begin{tabular}[c]{@{}l@{}}``[The data tracked by \\technology should not be] \\abused by insurance\\ companies'' (P04)\end{tabular}}  \\ \hline
\textbf{\begin{tabular}[c]{@{}l@{}}Sharing \\ Progress\end{tabular}} & \begin{tabular}[c]{@{}l@{}}FS \\ MO \\ SI\end{tabular} & \begin{tabular}[c]{@{}l@{}}Helpful for modifying exercises \\or proving that PT isn't\\ working; sharing can \\improve accountability\end{tabular} & \textit{\begin{tabular}[c]{@{}l@{}}``[sharing progress with PT] \\would help if a modification \\needed to be introduced\\ if something was severely\\ going wrong'' (P12)\end{tabular}}\\ \hline
\end{tabular}\\
\small 
TE = time and effort; FS = fluctuating symptoms; IN = insurance; MO = lack of motivation; SI = safety and injury
\end{table}

\subsubsection{Exercise Presentation}

Customization of how and what types of exercises were presented was important to participants so that they could account for fluctuating symptoms and decrease the time and effort required to complete their PT exercises.
Four participants wanted to customize their daily PT routine by choosing between multiple exercises in an application (P02, P04, P09, P11) so they could pick and choose according to their symptoms that day or equipment restrictions. 
As P04 explains, this option would be helpful \textit{``because certain exercises are targeted to help with certain areas...so only I would know what I need for the day.''} 
Three participants wanted different groups of exercises to be presented on different days (P10, P11, P12). 
Other participants had a subset of exercises that they would want to get prompted to do throughout the day such as exercises with minimal setup.
For example, P01 and P10 were  interested in doing sit-to-stands because \textit{``that's something you could do anywhere...even if you just had your wheelchair''} (P01).
Spreading out exercises that could be done with minimal setup throughout the day was appealing to participants because it alleviated the need to carve out time specifically to do the PT exercises. 

The other element that participants were most interested in customizing was the number of of repetitions for each exercise (P05, P10).
Customizing the number of repetitions could support participants in making an exercise goal more attainable.
For example, on a bad symptoms day, having a smaller goal may lead to more long-term success in adherence and goal completion than trying and failing to reach the same number of repetitions each day.

\rev{Manually accounting for fluctuating symptoms took extra time and effort and was a barrier to PT exercise completion.
An application that can decrease time and effort needed to make those daily adjustments around which exercises, what types of exercises, and the number of repetitions to complete can help alleviate such PT barriers.}

\subsubsection{Tracking}

Participants were excited to potentially use camera or sensor data to track at-home PT. They thought about tracking both individual movements (for quality and safety purposes) and overall progress (e.g., how many days did they do PT). If combined with \dcc~ symptom tracking, participants thought this data could support self experimentation and data-informed conversations with physical therapists or other medical care providers.

Tracking movement quality is key to ensure exercise quality and prevent injury. All but two (P05, P08) participants had one or more exercises they did at-home for which they wanted technology to provide feedback on movement quality. 
P04, P07, and P09 wanted feedback on movement quality to prevent injuries and pain, while others wanted to ensure they were doing the exercise correctly (P01, P02, P03, P06, P10-P12) and using the right muscle groups (P11). 
P10 was interested in leveraging exercise accuracy tracking to self-experiment to see how their movement quality changed over the course of the day as they got more tired. 

%
Participants were also excited to track PT progress so that they could keep track of and account for fluctuating symptoms. 
Six participants (P01, P04, P06, P09, P11-P12) discussed tracking PT progress to get \textit{``an actual physical visualization of what I'm able to do''} (P12). 
P06 suggested using the technology for self experimentation to better understand the efficacy of their PT; they commented that, if they could track their pain levels and the amount of PT exercises they are doing, they can see \textit{``if I stretch every day consistently for a month, does that decrease my pain levels or not.''}
In addition to supporting PT tracking, six participants (P01, P02, P06, P07-P09) were also interested in tracking other aspects of their life, such as tracking PT-like movements like walking during daily life or supporting tracking of \dcc~ symptoms such as pain, fatigue, or mental health, especially since symptoms and PT success and outcomes are so tightly intertwined.




Participants appreciated that better tracking encapsulated a more accurate representations of their daily practices into data that they could then share with their physical therapists. 
For example, four of our participants (P06, P08, P09, P11) were excited about being able to show more accurate pain levels and adherence to the PT routine over time. People felt that when their physical therapist asked them to rate pain levels in a session \textit{``it's just my answer at that time...if the number changes [later], I can't call them and tell them differently''} (P09). 
By tracking their PT progress and symptoms of \dcc, the participants felt that\textit{ ``the data can really help [the physical therapists] see daily [changes in \dcc~ symptoms]''} (P09) and it would be helpful to prove to physical therapists that \textit{``I am indeed doing ... what you've told me to''} (P08).

Two participants (P03, P12) thought that tracking exercise progression could be especially useful when in-person PT is not available to them: \textit{``I could actually keep using it even if I wasn't actively in [in-person] PT''} (P12).
In particular, adjusting the difficulty of the PT exercises between potentially infrequent PT visits can be challenging.
Tracking exercises could be helpful to support the unlocking of \textit{ ``new exercises that are similar to the ones that you've already been doing or if there's a way to build up the exercises that you're already doing...just make [the PT exercise] a little bit more challenging and a little bit more complex''} (P12).
Similarly, two participants wanted the technology to encourage progression by increasing the difficulty of the exercises or by increasing the number of repetitions required automatically (P03, P12).

\subsubsection{Technology Rewards}
Technology-based rewards could help keep PT engaging and motivating but participants differed in preferences as to what the rewards looked like.
Some participants (P02, P03, P04, P06, P08) were interested in unlocking a device (e.g., phone, laptop), while others (P01, P02, P04, P11) preferred to only gate access to less critical "time sink" activities (e.g., social media).
Using PT as a "commercial break" either before or as an interruption in a content consumption activity like watching TV, podcasts, or online videos was another popular suggestion (P02, P04, P07, P09, P10, P12). 
P12 mentioned how if they were \textit{``really in the thick of it with Grey's Anatomy, that would absolutely motivate me''}. 
Eight participants (P01, P03, P04, P06-P08, P10, P11) requested congratulatory or fun elements to be embedded into the technology to increase motivation.
P08 suggested a \textit{``sound notification....where it does a clap...or says `nice job on your walk'....[that] would make the user feel good about what they were doing''} and P04 mentioned \textit{``it should help me celebrate and reward when I'm keeping a good track...we like apps that make us feel good when we're doing something right''}.
P10 suggested a \textit{``wheel of PT''}, where a randomized feature can pick different exercises for them to do.
The final popular suggestion was having some type of game-like reward system, where they could accumulate points that add up to some sort of reward (P08, P09, P10, P11, P12). 
The rewards that were suggested varied, from free subscription services, to bloopers and special features for a show they are watching, and personalized quips and positive visual feedback for hitting certain goals.

\subsubsection{Notifications}
Participants saw the potential for personalized notifications to decrease the time and effort of remembering to do the exercises. 
Some participants preferred to spread out their exercises and notifications throughout the day (P03-P04, P08, P09, P11-P12), while others preferred to receive one notification and do all their exercises (P06-P07).
Additionally, P10 discussed the importance of the system customization extending to include notifications. For example, in addition to adjusting what and how many exercises to show on a  \textit{``bad pain day''}, the system could time notifications to be at low pain-points in the day.
Other participants saw the potential for more algorithmically driven notification timings based on their symptoms. 
For example, P09 was interested in receiving a PT exercise notification as a preemptive measure for or as a consequence of back pain:  
\textit{``I want it to detect my pain somehow and then prompt me''} (P09).

All participants defined the need to dismiss notifications when they were busy or having a bad symptoms day.
The most popular suggested dismissal method was a simple ``remind to do later'' option (P02, P04, P08, P11, P12). 
P03 suggested that, before fully dismissing, offering an easier exercise may encourage some PT rather than none in the moment. 
P04 and P12 additionally suggested disabling dismissals after a certain point to strongly encourage people to do the PT exercises or letting the person know that they have broken the record for the number of dismissals they have done. 
Finally, P12 suggested having the technology check-in with the person if they dismissed the notification too many times, which might indicate that a PT exercise is too challenging or that a person's symptoms are worsening.  

\subsubsection{Hardware Preferences}
As movement tracking and feedback was of interest, many participants were open to purchasing hardware such as cameras (e.g., \myrev{Wii U sensor bar}; P3, P07), wearables (e.g., Apple Watch; P04, P06, P09, P11), and implantable chips (e.g., Neuralink; P09). 
However, there were several concerns around the cost, upkeep, and comfort of hardware devices. For example, if the hardware device was a wearable, P02 mentioned \textit{``I'm not sure I would put it on everyday''} as a potential challenge, and P03 mentioned that \textit{``I'd be much, much less likely to use [wearables] ... because they would cause too much chafing,''} which is a concern for their \dcc. Others were concerned about both the base cost of hardware (P03, P05, P06-P07, P10-P11) as well as updates: \textit{``[hardware] would be a never-ending investment''}. Participants noted that insurance coverage could help ease the cost burden, but P03 was not optimistic: \textit{``I don't feel like insurance is ready for the video game technology yet''}. Without insurance support, many participants seemed reluctant to acquire extra hardware sensors for PT tracking.

\subsubsection{Data Security}

Insurance was a significant fiscal barrier to in-person PT access, and data security against insurance companies was a significant concern for tracking at-home PT.
For example, P05 commented that  \textit{``while movements on my arms might seem innocuous right now...''}, that information in the future could be used to prevent insurance coverage. 
P02-P04 were also concerned about the data being \textit{``abused by insurance companies''} (P03) because \textit{``the insurance company doesn't need to know that I missed a week of PT because I was hospitalized...the physical therapist understands that, and she isn't going to hold that against you, but the insurance company can and will''} (P04) because \textit{``...they already deny you for everything''} (P02). While tracking provides powerful affordances to support people in their PT experiences, technology designers must always take precautions to safeguard against this insurance abuse of their data.

\subsubsection{Sharing Progress}
All but two participants (P02, P05) were interested in sharing their at-home PT progress with their physical therapist to help with motivation, adjust exercises for fluctuating symptoms, and prevent injury. 
Seven participants (P01, P03, P04, P06, P09, P10, P12) thought that sharing progress with their physical therapist would be helpful \textit{``if a modification needed to be introduced''} (P12) and for the physical therapist \textit{``to have some control over how [the exercises] are programmed''} (P04). 
P06 felt that sharing the data would be helpful to work with their physical therapist to understand \textit{``where this PT regimen is working or it's not working''}.

%% file: PTaccess/sec_discussion.tex
\section{Discussion}
Clare writes about the tensions between \textit{``the wisdom that tells us the causes of the injustice we [disabled people] face lie outside our bodies, and also to the profound relationships our bodies have to that injustice''}~\cite{clare2001stolen}; the relationship between the body and built environment is evident in the experiences of people with \dcc~in performing PT. Our work highlights the tensions and intersections between social- and physiological-based barriers in PT access. 
For example, in-person PT can be inaccessible for people due to lack of public transportation (social barrier) because they can't drive due to their \dcc~ symptoms (physiological barrier).
While solutions like performing more at-home PT or sparsely attending in-person PT mitigate some of these issues, they introduce new access barriers. 
At-home PT can be difficult to complete if the person's \dcc~ symptoms are fluctuating or if there are concerns with safety and injury. 
Upon first glance, barriers like lack of transportation and fluctuating symptoms could be construed as individual barriers or compliance failures.
However, when viewed from an accessibility perspective, the locus of responsibility for addressing such barriers may shift (such as reconsidering the accessibility of the transportation system as a social responsibility) as can the strategy for addressing challenges (such as expanding technology from monitoring to addressing fluctuating symptoms). 
%
In the words of P11, \textit{``I'm just overwhelmed, and if technology can help make sort of a systematic way to address chronic illness challenges and hopefully the medical system will follow suit, I think that would be a huge benefit to people like me''}.


\subsection{Design Recommendations to Support PT With Technology}

People with \dcc~who are doing PT inherently face physiologically-based barriers to PT access: symptoms of the chronic condition that they are doing PT for. 
Although a traditional medical interpretation of such physiological barriers to PT access suggests a physiologically-based solution (e.g., improve motivation and engagement through tracking and gamification of the PT exercises), considering both physiological and social access barriers, and how such barriers can be reinterpreted to focus on the individual's lived experience is important for developing technology that holistically supports the person's PT goals.
We present three design recommendations (adaptability, movement tracking, community building) and one tension (insurance) that highlights the nuanced interplay between social and physiological access barriers and how technology solutions to PT access barriers can be reconsidered to take into account the lived experience of people with \dcc. 
Although we do not specifically call out accessibility as a design recommendation, we emphasize that first and foremost, any PT technology must be built with digital accessibility (e.g., screen reader accessibility) in mind from the start to prevent further inaccess to PT~\cite{rinne2016democratizing}. 
Technologies that continuously adapt to the person's abilities~\cite{alankus2015reducing} and enable accessible inputs and calibration~\cite{alankus2010towards,huang2014technology} are ideal.



\subsubsection{Recommendation: Design with Adaptability in Mind} 
Perhaps the most complex and pervasive issues mentioned in our interviews focused on the physiological barriers that arose between PT and a person's body. 
These barriers included feeling too symptomatic to perform exercises, exercises triggering symptoms that persist throughout the day, fluctuating symptoms, and complex interactions with multi-faceted and/or multiple conditions. 
Such barriers are traditionally considered ``adherence issues'', both externalized by physical therapists who encourage adherence by encouraging the individual doing PT to remember their larger goals for doing PT~\cite{huang2014technology} and internalized by the individuals themselves, who can have feelings of guilt when they do not adhere to the exercise routines. 

However, if we adopt an accessibility lens, we can view the lack of adherence as not an physiological issue, but one that arises from a mismatch between the prescribed PT exercises and how the individual is feeling on that day. 
This mismatch can be especially potent for people who have multiple conditions. 
For example, an exercise that alleviates one impairment or symptom may exacerbate another, or an individual is tasked with doing an insurmountable number of PT exercises for each of their multiple conditions.
These complex, contradictory health needs could be alleviated with attentive care and guidance from a physical therapist that can adapt exercises to match people's abilities for that day or holistically evaluate a person's multiple conditions.
\myrev{
This barrier can be compounded by social barriers like ableism or medical bias~\cite{kirschner2009educating,iezzoni2021physicians}, resulting in the physical therapist prescribing exercises that are not accessible to the individual doing PT~\cite{rinne2016democratizing}.
Designing therapy through a collaborative and iterative process has previously been shown to be critical to support individuals with chronic pain and other disabilities~\cite{vowles2012patient,moore2018framework,orlin2014continuum}. 
Other social barriers like insurance can also prevent people from accessing in-person PT or the structure of the medical system can make it challenging for a person to get a holistic evaluation and exercise recommendations for their multiple conditions.
}


In such scenarios, the individual doing the PT is the expert in their own condition.
Technology that supports people with \dcc~doing PT should ensure that people can flexibly adapt the exercises and the number of repetitions to match what they need for their body on that day and provide people with the data they need to make those decisions. 
For example, applications to support PT should be built in a way that records the abilities (e.g., pain level less than 4/10, dizziness less than 3/10) and time needed to complete the exercise. 
Then, when a user chooses to perform PT, the exercises and the number of repetitions can be tailored based on the individual's symptoms and abilities to increase the odds of success. 
To fully and holistically support people in managing their \dcc, adaptability must be built-in to technology supports.

\subsubsection{Recommendation: Allow for Tracking Movement Quality and Progress}

Due to recent advances in wearable sensors and decreased computing cost, technology is opportunistically situated to enable tracking for movement quality and exercise progress.
Tracking movement quality was a highly requested feature that our participants felt would help ensure safety and prevent injury.
Tracking exercises to prevent injury using wearables or camera-based technology is currently being explored in literature~\cite{huang2014technology}, but not necessarily within a PT context~\cite{rector2013eyes}. 
With the ability to track the quality of PT exercises, at-home PT becomes more accessible to people with \dcc~who are concerned about the potential for muscle or joint injury without physical therapist supervision. 

Additionally, technology can reduce the burden of tracking exercise progress and completion to highlight people's successes and improve motivation, aid in self-experimentation and adaptation of PT exercises, and open the door to 
\myrev{making decisions and advocating for their needs by}
having conversations with physical therapists about adapting exercises.  
Self-tracking can be burdensome and another way that managing a medical condition can shorten the day for people with \dcc~\cite{epstein2015lived, vizer2019s}.
To alleviate this time-burden, wearable sensors or camera-based technologies can be used to automatically track the type and number of exercises completed. 
Algorithmic methods that enable safe and automated tracking of PT exercises is an unsolved issue that requires further research.
During algorithm development, it is important to consider our participants' concerns about purchasing additional hardware.
Therefore, supporting tracking with wearable technologies that people may already own such as smartwatches or enabling tracking with solely software-based solutions using a smartphone may make it easier for people to incorporate automated tracking into their PT exercise routines.


\subsubsection{Recommendation: Create Technology That Encourages Community-Building}

Our participants felt that being accountable to a community, whether it be with other people doing PT or with a physical therapist, is helpful for staying motivated to complete their exercises.
\rev{In one study of a community-building PT application, participants found that the community was helpful for improving motivation and for comparing their PT exercises to other people who had similar conditions so they could experiment with new PT exercises~\cite{malu2017sharing}. 
Although there were concerns with misleading information~\cite{malu2017sharing}, information sharing could be a useful work-around for when people are unable to see a physical therapist to get updated exercises. 
Additionally, virtual communities could facilitate encouragement and engagement between users, especially those who have rare or multiple conditions~\cite{malu2017sharing}. }

Facilitating supportive interactions between the person and their physical therapist through sharing of PT progress between in-person sessions may also increase familiarity and a sense of accomplishment between the individual and the physical therapist.
It can be challenging for physical therapists to keep track of how individuals are doing in-between sessions~\cite{huang2014technology}.
Consistent data-sharing could help catch issues with the PT routines in a timely manner and prevent communication issues.
In situations where people are doing less than the prescribed number of exercises, it is important to frame the incomplete adherence as a sign that the PT routine is too time consuming, hard, or challenging because of other symptoms, rather than simply an adherence issue, to prioritize the lived experiences of the person doing PT.
This reframing highlights where the PT routine might not be the best suited for the person, not that the person is primarily or solely to blame for non-adherence. 
These discussions, supported by data tracking, will hopefully lead to more appropriate exercise selection given a more holistic view of the person and lead to increased PT success, and therefore, adherence.

\subsubsection{Tension: Designing with Insurance in Mind}

Finally, to support tracking of PT with technology, we cannot ignore the constant presence of insurance concerns. Insurance was a major barrier to PT access for many participants, all of whom were located in the US. While technology could help participants work around the in-person limits insurance imposes, participants were concerned about the ability for data collected with technology to be used against them. Therefore, technology that is built to support PT must be built with data privacy and use in mind.

Difficulties with and mistrust of insurance were near-ubiquitous in our interviews. 
The first major barrier to PT that many people faced was lack of coverage. 
Generally, one needs to be employed to receive benefits that include PT coverage in the US, which already excludes a considerable number of disabled people from receiving PT.
Even if someone with a \dcc~ wanted more flexible work hours or less working hours to better manage their symptoms, that is often not feasible because benefits are directly tied to the job. 
While health insurance is available in the US for unemployed individuals, navigating that infrastructure with a \dcc~ can be challenging, and the coverage can be insufficient\rev{~\cite{rogers2018medicaid,patterson2013access}.}
One major improvement that could be made in this area is to deploy (accessible) technology to help navigate finding and obtaining insurance that provides sufficient coverage for people's needs. 


Further, these problems around achieving and maintaining PT coverage were so pervasive in our participants' lives that it affected their willingness to use technology to support PT. While a few participants thought a PT system that tracks progress and movements could serve as evidence for skeptical doctors or physical therapists, many other participants were uncomfortable with the PT app tracking progress because of the potential insurance ramifications; they were concerned that their progress, or lack thereof, could be acquired by insurance companies and used to deny coverage. Therefore, though technology can be part of the solution for increasing the accessibility of PT, insurance politics indubitably need to change before the access challenges are fully removed.

With these insights in mind, technology that supports PT must consider data privacy. 
\rev{Most work on data privacy in healthcare focuses on securing medical data collected by the healthcare industry against security breaches~\cite{patil2014big,abouelmehdi2017big}
Our results bring up intriguing considerations for data privacy policies and systems to protect private patient data from being misused by the healthcare system, including insurance companies.}
Further, our work uncovers complex ethical considerations when creating such technology. If technology that could have negative consequences for users is built and adopted by insurance companies as ``required for coverage,'' the technology we use and the data it collects could actively harm participants with respect to PT and healthcare access.


\subsection{Limitations and Future Work}

One area of PT that we did not discuss in this paper is the rising popularity of telemedicine or virtual PT, where the physical therapist and the individual receiving care meet on video call for diagnosis and to receive care~\cite{demartini2020physical,schneider2017promise}. 
Similar to reflections on virtual workplaces \cite{Linden14, Moon14, McNaughton14}, a new, virtual setting for PT can improve some accessibility issues while introducing new challenges. 
Telemedicine can alleviate some aspects of the physical inaccessibility of in-person PT, such as removing the need to find transportation and the risks of exposure for people with weakened immune systems. 
However, similar barriers to at-home PT such as the potential for injury, lack of equipment, and lack of safety without the physical presence of the physical therapists exist in the virtual setting~\cite{turolla2020musculoskeletal}. 
Additionally, virtual PT can be difficult because video calling can only provide so much information to properly diagnose and provide recommendations to the person doing the PT exercises~\cite{turolla2020musculoskeletal}.
Here, new and emerging sensors such as wearables and 3D cameras could be leveraged to provide the physical therapist with the information they need to care for their clients.

Additionally, PT access is not only challenging for people with chronic health conditions. 
People who are blind or low vision, d/Deaf or hard-of-hearing, and/or neurodiverse could also need PT access for acute (e.g., breaking a leg) or chronic (e.g., having a stroke) conditions. 
Consequently, the accessibility of non-visual PT, for example, is a key part of injury recovery for people who are blind or low vision, but it is not the focus of current accessibility research or standards.
Although technology can never replace the knowledge and expertise of physical therapists, our study begins to identify potential ways in which technology can improve PT access, particularly if technologies are designed with access in mind from the start.

\rev{In the broader context of our conversations with our participants, many of the social and physiological PT access barriers that participants mentioned were barriers to healthcare access in general.
Yet, accessibility research in healthcare remains limited. 
Our work highlights the many areas (reminders, tracking, motivational rewards, data security) where technology could improve healthcare access for people with \dcc. }

Lastly, beyond PT access, our study revealed several HCI areas in which needs arose at the intersection of social and physiological access barriers for people with \dcc. 
For example, mental health was a PT barrier for many participants and is generally an understudied chronic condition in HCI accessibility literature \cite{mack2021what} that limits participation in daily life due to both social and physiological barriers~\cite{world2001world}. 
Further, almost all of our participants discussed overlapping social and physiological access barriers to performing their jobs, such as having to limit work hours due to chronic pain. 
We also suggest taking a post-modern lens of disability to other important situations for people with \dcc, such as education, leisure activities, or care work. 
Given the pervasiveness of access barrier stemming from the body and society in many areas of life for people with chronic conditions, we see the intersection of physiological and social access barriers as a rich area of future HCI accessibility research.